\documentstyle[amssymb,aps,12pt]{revtex}

\begin{document}
\draft
\author{Sergio De Filippo}
\address{Dipartimento di Fisica ''E.R. Caianiello'', Universit\`{a} di Salerno\\
Via Allende I-84081 Baronissi (SA) ITALY\\
Tel: +39 089 965 229, Fax: +39 089 965 275, e-mail: defilippo@sa.infn.it\\
and \\
Unit\`{a} I.N.F.M., I.N.F.N. Salerno}
\date{\today}
\title{Non-unitary HD gravity classically equivalent to Einstein gravity}
\maketitle

\begin{abstract}
Runaway solutions can be avoided in fourth order gravity by a doubling of
the matter operator algebra with a symmetry constraint with respect to the
exchange of observable and hidden degrees of freedom together with the
change in sign of the ghost and the dilaton fields. The theory is
classically equivalent to Einstein gravity, while its non-unitary Newtonian
limit is compatible with the wavelike properties of microscopic particles
and the classical behavior of macroscopic bodies, as well as with a
trans-Planckian regularization of collapse singularities. A unified reading
of ordinary and black hole entropy emerges as entanglement entropy with
hidden degrees of freedom.
\end{abstract}


Although higher derivative (HD) gravity has long been popular as a natural
generalization of Einstein gravity\cite{dewitt,stelle,nojiri}, since
''perturbation theory for gravity ... requires higher derivatives in the
free action''\cite{hawkinghertog}, already on the classical level it is
unstable due to negative energy fields giving rise to runaway solutions\cite
{hawkinghertog}. On the quantum level an optimistic conclusion as to
unitarity is that ''the S-matrix will be nearly unitary\cite{dewitt}''\cite
{hawkinghertog}.

A way out of the so called information loss paradox\cite{hawking1,preskill}
of black hole physics\cite{hawking2} may be precisely a fundamental
non-unitarity\cite{ellis,banks,unruh,kay,wald}: ''For almost any initial
quantum state, one would expect ... a nonvanishing probability for evolution
from pure states to mixed states''\cite{unruh}. Though such an evolution is
incompatible with a cherished principle of quantum theory, the crucial issue
is to see if it necessarily gives rise to a loss of coherence or to
violations of energy-momentum conservation so large as to be incompatible
with ordinary laboratory physics \cite{ellis,banks,unruh,wald}, as guessed
for Markovian effective evolution laws \cite{ellis,banks}. However one
expects that a law modeling black hole formation and evaporation, far from
being local in time, should retain a long term ``memory''\cite{unruh,wald}.

Here a specific non-unitary realization of HD gravity is shown to be
classically stable, as well as compatible with the wavelike properties of
microscopic particles and with the assumption of a gravity-induced emergence
of classicality\cite
{karolyhazy,hawking3,diosi,ghirardi,penrose,squires,ellis2,anandan}.
Moreover it leads to the reading of the thermodynamical entropy of a closed
system as von Neumann entropy, or equivalently as entanglement entropy with
hidden degrees of freedom\cite{unruh,wald}, which allows, in principle, to
overcome the dualistic nature of the notions of ordinary and
Bekenstein-Hawking (B-H) entropy \cite{bekenstein}. To be specific, the B-H
entropy\cite{bekenstein} may be identified with the von Neumann entropy of
the collapsed matter, or equivalently with the entanglement entropy between
matter and hidden degrees of freedom, both close to the smoothed
singularity. In fact the model seems to give clues for the elimination of
singularities on a trans-Planckian scale. Parenthetically we are encouraged
in our extrapolations by the success of inflationary models, implicitly
referring to these scales\cite{brandenberger}. This reading of B-H entropy
may appear rather natural, as the high curvature region is where new physics
is likely to emerge. However, in passing from the horizon\cite{wald}, where
quantum field theory in curved space-times is expected to work, to the
region close to the classical singularity, in the absence of a full theory
of quantum gravity, we have to rely on heuristic arguments and some guessing
work, which we intend to show can be carried out by rather natural
assumptions. This reading, however, is corroborated by the attractive
features of the Newtonian limit of the model.

Similarly to Ref. \cite{hawkinghertog}, we consider first a simpler fourth
order theory for a scalar field $\phi $, which has the same ghostly behavior
as HD gravity. Its action 
\begin{equation}
S=\int d^{4}x\left[ -\phi \square \left( \square -\mu ^{2}\right) \phi
/2-\lambda \phi ^{4}+\alpha \psi ^{\dagger }\psi \phi \right] +S_{mat}\left[
\psi ^{\dagger },\psi \right]
\end{equation}
includes a matter action $S_{mat}$ and an interaction with matter, where $%
\psi ^{\dagger }\psi $ is a shorthand notation for a quadratic scalar
expression in matter fields. Defining 
\begin{equation}
\phi _{1}=\left( \square -\mu ^{2}\right) \phi /\mu ,\;\phi _{2}=\square
\phi /\mu ,\;  \label{transformation}
\end{equation}
the action can be rewritten as 
\begin{eqnarray}
&&S\left[ \phi _{1},\phi _{2},\psi ^{\dagger },\psi \right]  \nonumber \\
&=&\int d^{4}x\left[ \frac{1}{2}\phi _{1}\square \phi _{1}-\frac{1}{2}\phi
_{2}\left( \square -\mu ^{2}\right) \phi _{2}-\lambda \left( \frac{\phi
_{2}-\phi _{1}}{\mu }\right) ^{4}+\frac{\alpha }{\mu }\psi ^{\dagger }\psi
\left( \phi _{2}-\phi _{1}\right) \right] +S_{mat}\left[ \psi ^{\dagger
},\psi \right] .  \label{secorder}
\end{eqnarray}
The quadratic term in $\phi _{2}$ has the wrong sign, which classically
means that the energy of this field is negative. Due to the presence of
interactions, energy can flow from negative to positive energy degrees of
freedom, and one can have runaway solutions\cite{hawkinghertog}.

In this model there is a cancellation of all self-energy and vertex
infinities coming from the $\psi ^{\dagger }\psi \phi $ interaction, owing
to the difference in sign between $\phi _{1}$ and $\phi _{2}$ propagators.
This feature, ''analogous to the Pauli-Villars regularization of other field
theories'' \cite{stelle}, is responsible for the improved ultraviolet
behavior in HD gravity \cite{stelle}. A key feature of the non-interacting
theory ($\lambda =\alpha =0$), making it classically viable, can be
considered to be its symmetry under the transformation $\phi
_{2}\longrightarrow -\phi _{2}$, by which symmetrical initial conditions
with $\phi _{2}=0$ produce symmetrical solutions. If one symmetrizes the
Lagrangian (\ref{secorder}) as it is, in order to extend this symmetry to
the interacting theory, this eliminates the interaction between the ghost
field and the matter altogether and, with it, the mentioned cancellations. A
possible procedure to get a symmetric action while keeping cancellations is
suggested by previous attempts \cite{hawkinghertog} and by the information
loss paradox \cite{unruh}, both pointing to a non-unitary theory with hidden
degrees of freedom. In particular the most natural way to make the hidden
degrees of freedom ''not ... available as either a net source or a sink of
energy'' \cite{unruh} is to constraint them to be a copy of observable ones.
Accordingly we introduce a (meta-)matter algebra that is the product of two
copies of the observable matter algebra, respectively generated by the $\psi
^{\dagger },\psi $ and $\tilde{\psi}^{\dagger },\tilde{\psi}$ operators, and
a symmetrized action 
\begin{equation}
S_{Sym}=\left\{ S\left[ \phi _{1},\phi _{2},\psi ^{\dagger },\psi \right] +S%
\left[ \phi _{1},-\phi _{2},\tilde{\psi}^{\dagger },\tilde{\psi}\right]
\right\} /2,  \label{symac}
\end{equation}
which is invariant under the symmetry transformation 
\begin{equation}
\phi _{1}\longrightarrow \phi _{1},\;\;\phi _{2}\longrightarrow -\phi
_{2},\;\;\psi \longrightarrow \tilde{\psi},\;\;\tilde{\psi}\longrightarrow
\psi .\;
\end{equation}

If the symmetry constraint is imposed on states $\left| \Psi \right\rangle $%
, i.e. the state space is restricted to those states that are generated from
the vacuum by symmetrical operators, then 
\begin{equation}
\left\langle \Psi \right| F\left[ \phi _{2},\psi ^{\dagger },\psi \right]
\left| \Psi \right\rangle =\left\langle \Psi \right| F\left[ -\phi _{2},%
\tilde{\psi}^{\dagger },\tilde{\psi}\right] \left| \Psi \right\rangle
\;\;\forall F.  \label{constraint}
\end{equation}
The allowed states do not give a faithful representation of the original
algebra, which is then larger than the observable algebra. In particular
they cannot distinguish between $F\left[ \psi ^{\dagger },\psi \right] $ and 
$F\left[ \tilde{\psi}^{\dagger },\tilde{\psi}\right] $, by which the $\tilde{%
\psi}$ operators are referred to hidden degrees of freedom\cite{unruh}. On a
classical level $\psi $ and $\tilde{\psi}$ are identified, the $\phi _{2}$
field vanishes and the classical constrained action is that of an ordinary
second order scalar theory interacting with matter: 
\begin{equation}
S_{Cl}=\int d^{4}x\left[ \phi _{1}\square \phi _{1}/2-\lambda \left( \phi
_{1}/\mu \right) ^{4}-\alpha \phi _{1}\psi ^{\dagger }\psi /\mu \right]
+S_{mat}\left[ \psi ^{\dagger },\psi \right] .
\end{equation}
\qquad \qquad 

Consider now the action of a fourth order theory of gravity including matter 
\cite{stelle}

\begin{eqnarray}
S &=&S_{G}\left[ g_{\mu \nu }\right] +S_{mat}\left[ g_{\mu \nu },\psi
^{\dagger },\psi \right]  \nonumber \\
&=&-\int d^{4}x\sqrt{-g}\left[ \alpha R_{\mu \nu }R^{\mu \nu }-\beta
R^{2}+R/\left( 16\pi G\right) \right] +\int d^{4}x\sqrt{-g}L_{mat},
\label{hd}
\end{eqnarray}
where $L_{mat}$ denotes the matter Lagrangian density in a generally
covariant form. In terms of the contravariant metric density 
\begin{equation}
\sqrt{32\pi G}h^{\mu \nu }=\sqrt{-g}g^{\mu \nu }-\eta ^{\mu \nu },
\end{equation}
the Newtonian limit of the static field is 
\begin{equation}
h^{00}\sim 1/r+e^{-\mu _{0}r}/(3r)-4e^{-\mu _{2}r}/(3r),  \label{potential}
\end{equation}
where $\mu _{0}=[32\pi G(3\beta -\alpha )]^{-1/2}$, $\mu _{2}=[16\pi G\alpha
]^{-1/2}$\cite{stelle}. From Stelle's linearized analysis\cite{stelle}, the
first term in Eq. (\ref{potential}) corresponds to the graviton, the second
one to a massive scalar and the third one to a negative energy spin-two
field. In fact, in analogy with Eq. (\ref{transformation}), one can
introduce a transformation from $g_{\mu \nu }$ to a new metric tensor $\bar{g%
}_{\mu \nu }$, a scalar field $\chi $ dilatonically coupled to $\bar{g}_{\mu
\nu }$ and a spin-two field $\phi _{\mu \nu }$, this transformation leading
to the second order form of the action\cite{hindawi}. To be specific,
referring to Ref.\cite{hindawi} (see Eq. (6.9) apart from the matter term),
the action (\ref{hd}) becomes the sum of the Einstein-Hilbert action $S_{EH}$
for $\bar{g}_{\mu \nu }$, an action $S_{gh}$ for $\phi _{\mu \nu }$ and $%
\chi $ coupled to $\bar{g}_{\mu \nu }$, and a matter action $S_{mat}$, with $%
g_{\mu \nu }$ expressed in terms of $\bar{g}_{\mu \nu }$, $\phi _{\mu \nu }$
and $\chi $ (replacing $g_{\mu \nu }$ by $e^{\chi }g_{\mu \nu }$ in Eq.
(4.12) in Ref. \cite{hindawi}): 
\begin{eqnarray}
&&S\left[ \bar{g}_{\mu \nu },\phi _{\mu \nu },\chi ,\psi ^{\dagger },\psi %
\right]  \nonumber \\
&=&S_{EH}\left[ \bar{g}_{\mu \nu }\right] +S_{gh}\left[ \bar{g}_{\mu \nu
},\phi _{\mu \nu },\chi \right] +S_{mat}\left[ g_{\mu \nu }(\bar{g}_{\sigma
\tau },\phi _{\sigma \tau },\chi ),\psi ^{\dagger },\psi \right] .
\label{secordergr}
\end{eqnarray}

In $S_{gh}$ the quadratic part in $\phi _{\mu \nu }$ has the wrong sign \cite
{hindawi}. One could symmetrize the action with respect to the
transformation $\phi _{\mu \nu }\rightarrow -\phi _{\mu \nu }$, but this
would eliminate the repulsive term in Eq. (\ref{potential}), which below
plays a role in avoiding the singularity in gravitational collapse. Like for
the toy model, we double the matter algebra and define the symmetrized
action 
\begin{equation}
S_{Sym}=\left\{ S\left[ \bar{g}_{\mu \nu },\phi _{\mu \nu },\chi ,\psi
^{\dagger },\psi \right] +S\left[ \bar{g}_{\mu \nu },-\phi _{\mu \nu },-\chi
,\tilde{\psi}^{\dagger },\tilde{\psi}\right] \right\} /2,  \label{covsymac}
\end{equation}
which is symmetric under the transformation 
\begin{equation}
\bar{g}_{\mu \nu }\longrightarrow \bar{g}_{\mu \nu },\;\;\phi _{\sigma \tau
}\rightarrow -\phi _{\sigma \tau },\;\chi \longrightarrow -\chi ,\;\psi
\longrightarrow \tilde{\psi},\;\tilde{\psi}\longrightarrow \psi .\;
\end{equation}
If only symmetric states are allowed, the $\tilde{\psi}$ operators denote
hidden degrees of freedom, as 
\begin{equation}
\left\langle \Psi \right| F\left[ \bar{g}_{\mu \nu },\phi _{\mu \nu },\chi
,\psi ^{\dagger },\psi \right] \left| \Psi \right\rangle =\left\langle \Psi
\right| F\left[ \bar{g}_{\mu \nu },-\phi _{\mu \nu },-\chi ,\tilde{\psi}%
^{\dagger },\tilde{\psi}\right] \left| \Psi \right\rangle \;\;\forall F.
\end{equation}
On a classical level $\psi $ and $\tilde{\psi}$ are identified, while the $%
\phi _{\mu \nu }$ and $\chi $ fields vanish and the classical constrained
action is that of \ ordinary matter coupled to ordinary gravity: 
\begin{equation}
S_{Cl}\left[ \bar{g}_{\mu \nu },\psi ^{\dagger },\psi \right] =S_{EH}\left[ 
\bar{g}_{\mu \nu }\right] +S_{mat}\left[ \bar{g}_{\mu \nu },\psi ^{\dagger
},\psi \right] ,  \label{einstein}
\end{equation}
as $S_{gh}\left[ \bar{g}_{\mu \nu },0,0\right] =0$ (Eq. (6.9) in Ref. \cite
{hindawi}) and $g_{\mu \nu }(\bar{g}_{\sigma \tau },0,0)=\bar{g}_{\sigma
\tau }$ (Eq. (4.12) in Ref. \cite{hindawi} with $e^{\chi }g_{\mu \nu }$
replacing $g_{\mu \nu }$).

After the elimination of classical runaway solutions, a further natural step
in assessing the consistency of the theory is the study of its implications
for ordinary laboratory physics. Consider the Newtonian limit with
non-relativistic meta-matter and instantaneous action at a distance. By Eq. (%
\ref{covsymac}), we see that the interactions due to $\bar{g}_{\mu \nu }$
are always attractive, whereas those due to $\phi _{\mu \nu }$ are repulsive
within observable and within hidden meta-matter, as shown by the minus sign
in Eq. (\ref{potential}), and are otherwise attractive, as the ghostly
character is offset by the difference in sign in its coupling with
observable and hidden meta-matter; the reverse is true for the scalar field $%
\chi $. The corresponding (meta-)Hamiltonian is 
\begin{eqnarray}
H_{G} &=&H_{0}[\psi ^{\dagger },\psi ]-\frac{G}{4}\sum_{j,k}m_{j}m_{k}\int
dxdy\frac{:\psi _{j}^{\dagger }(x)\psi _{j}(x)\psi _{k}^{\dagger }(y)\psi
_{k}(y):}{|x-y|}\left( 1+\frac{e^{-\mu _{0}\left| x-y\right| }}{3}-\frac{%
4e^{-\mu _{2}\left| x-y\right| }}{3}\right)   \nonumber \\
&&+H_{0}[\tilde{\psi}^{\dagger },\tilde{\psi}]-\frac{G}{4}%
\sum_{j,k}m_{j}m_{k}\int dxdy\frac{:\tilde{\psi}_{j}^{\dagger }(x)\tilde{\psi%
}_{j}(x)\tilde{\psi}_{k}^{\dagger }(y)\tilde{\psi}_{k}(y):}{|x-y|}\left( 1+%
\frac{e^{-\mu _{0}\left| x-y\right| }}{3}-\frac{4e^{-\mu _{2}\left|
x-y\right| }}{3}\right)   \nonumber \\
&&-\frac{G}{2}\sum_{j,k}m_{j}m_{k}\int dxdy\frac{\psi _{j}^{\dagger }(x)\psi
_{j}(x)\tilde{\psi}_{k}^{\dagger }(y)\tilde{\psi}_{k}(y)}{|x-y|}\left( 1-%
\frac{e^{-\mu _{0}\left| x-y\right| }}{3}+\frac{4e^{-\mu _{2}\left|
x-y\right| }}{3}\right)   \label{metahamiltonian}
\end{eqnarray}
acting on the product $F_{\psi }\otimes F_{\tilde{\psi}}$ of the Fock spaces
of (the non-relativistic counterparts of) $\psi $ and $\tilde{\psi}$. Two
couples of meta-matter operators $\psi _{j}^{\dagger },\psi _{j}$ and $%
\tilde{\psi}_{j}^{\dagger },\tilde{\psi}_{j}$ \ appear for every particle
species and spin component, while $m_{j}$ is the mass of the $j$-th species
and $H_{0}$ is the gravitationless matter Hamiltonian. The $\tilde{\psi}$
operators obey the same statistics as the corresponding operators $\psi $,
while $[\psi ,\tilde{\psi}]$ $_{-}=[\psi ,\tilde{\psi}^{\dagger }]_{-}=0$.
Tracing out $\tilde{\psi}$ from a symmetrical meta-state evolving according
to the unitary meta-dynamics generated by $H_{G}$ results in a non-Markov
non-unitary physical dynamics for the ordinary matter algebra\cite
{defilippo1}.

Considering, for simplicity, particles of one and the same species, the time
derivative of the matter canonical momentum in a space region $\Omega $ in
the Heisenberg picture reads 
\begin{eqnarray}
\frac{d\overrightarrow{p}_{\Omega }}{dt} &=&-i\hslash \frac{d}{dt}%
\int_{\Omega }dx\psi ^{\dagger }(x)\nabla \psi (x)\equiv \left. \frac{d%
\overrightarrow{p}_{\Omega }}{dt}\right| _{G=0}+\vec{F}_{G}=-\frac{i}{%
\hslash }\left[ \overrightarrow{p}_{\Omega },H_{0}[\psi ^{\dagger },\psi ]%
\right]  \nonumber \\
&&+\frac{G}{2}m^{2}\int_{\Omega }dx\psi ^{\dagger }(x)\psi (x)\nabla
_{x}\int_{R^{3}}dy\frac{\tilde{\psi}^{\dagger }(y)\tilde{\psi}(y)}{\left|
x-y\right| }\left( 1-\frac{e^{-\mu _{0}\left| x-y\right| }}{3}+\frac{%
4e^{-\mu _{2}\left| x-y\right| }}{3}\right)  \nonumber \\
&&+\frac{G}{2}m^{2}:\int_{\Omega }dx\psi ^{\dagger }(x)\psi (x)\nabla
_{x}\int_{R^{3}}dy\frac{\psi ^{\dagger }(y)\psi (y)}{\left| x-y\right| }%
\left( 1+\frac{e^{-\mu _{0}\left| x-y\right| }}{3}-\frac{4e^{-\mu _{2}\left|
x-y\right| }}{3}\right) :.
\end{eqnarray}
The expectation of the gravitational force can be written as 
\begin{eqnarray}
\left\langle \vec{F}_{G}\right\rangle &=&\frac{G}{2}m^{2}\left\langle
\int_{\Omega }dx\psi ^{\dagger }(x)\psi (x)\nabla _{x}\int_{\Omega }dy\frac{%
\tilde{\psi}^{\dagger }(y)\tilde{\psi}(y)}{\left| x-y\right| }\left( 1-\frac{%
e^{-\mu _{0}\left| x-y\right| }}{3}+\frac{4e^{-\mu _{2}\left| x-y\right| }}{3%
}\right) \right\rangle  \nonumber \\
&&+\frac{G}{2}m^{2}\left\langle \int_{\Omega }dx\psi ^{\dagger }(x)\psi
(x)\nabla _{x}\int_{R^{3}\backslash \Omega }dy\frac{\tilde{\psi}^{\dagger
}(y)\tilde{\psi}(y)}{\left| x-y\right| }\left( 1-\frac{e^{-\mu _{0}\left|
x-y\right| }}{3}+\frac{4e^{-\mu _{2}\left| x-y\right| }}{3}\right)
\right\rangle  \nonumber \\
&&+\frac{G}{2}m^{2}\left\langle :\int_{\Omega }dx\psi ^{\dagger }(x)\psi
(x)\nabla _{x}\int_{\Omega }dy\frac{\psi ^{\dagger }(y)\psi (y)}{\left|
x-y\right| }\left( 1+\frac{e^{-\mu _{0}\left| x-y\right| }}{3}-\frac{%
4e^{-\mu _{2}\left| x-y\right| }}{3}\right) :\right\rangle  \nonumber \\
&&+\frac{G}{2}m^{2}\left\langle \int_{\Omega }dx\psi ^{\dagger }(x)\psi
(x)\nabla _{x}\int_{R^{3}\backslash \Omega }dy\frac{\psi ^{\dagger }(y)\psi
(y)}{\left| x-y\right| }\left( 1+\frac{e^{-\mu _{0}\left| x-y\right| }}{3}-%
\frac{4e^{-\mu _{2}\left| x-y\right| }}{3}\right) \right\rangle ,
\end{eqnarray}
where, on allowed states, the first term vanishes for the antisymmetry of
the kernel $\nabla _{x}\left[ \left( 1-e^{-\mu _{0}|x-y|}/3+4e^{-\mu
_{2}|x-y|}/3\right) /\left| x-y\right| \right] $ and the symmetry of the
state, while the third one vanishes just as a consequence of the
antisymmetry of the corresponding kernel. We can approximate $\left\langle
\psi ^{\dagger }(x)\psi (x)\tilde{\psi}^{\dagger }(y)\tilde{\psi}%
(y)\right\rangle $ and $\left\langle \psi ^{\dagger }(x)\psi (x)\psi
^{\dagger }(y)\psi (y)\right\rangle $ respectively with $\left\langle \psi
^{\dagger }(x)\psi (x)\right\rangle \left\langle \tilde{\psi}^{\dagger }(y)%
\tilde{\psi}(y)\right\rangle $ and $\left\langle \psi ^{\dagger }(x)\psi
(x)\right\rangle \left\langle \psi ^{\dagger }(y)\psi (y)\right\rangle $ ,
as $x\in \Omega $ and $y\in R^{3}\backslash \Omega $. Finally, as $%
\left\langle \tilde{\psi}^{\dagger }(y)\tilde{\psi}(y)\right\rangle
=\left\langle \psi ^{\dagger }(y)\psi (y)\right\rangle $, we get 
\begin{equation}
\left\langle \vec{F}_{G}\right\rangle \simeq Gm^{2}\int_{\Omega
}dx\left\langle \psi ^{\dagger }(x)\psi (x)\right\rangle \nabla
_{x}\int_{R^{3}\backslash \Omega }dy\left\langle \psi ^{\dagger }(y)\psi
(y)\right\rangle /\left| x-y\right| ,
\end{equation}
as for the traditional Newton interaction between observable degrees of
freedom only, consistently with the classical equivalence of the original
theory to Einstein gravity.

As to the quantum aspects of the present Newtonian model, a closely related
model was analyzed in Ref.s \cite{defilippo1}. Actually, if in Ref. \cite
{defilippo1} $H[\psi ^{\dagger },\psi ]$ is meant to be the sum of $%
H_{0}[\psi ^{\dagger },\psi ]$ and the normal ordered interaction within
observable matter in Eq. (\ref{metahamiltonian}) above, and analogously for
the hidden meta-matter, the two models differ only for the kernel in the
interaction between observable and hidden meta-matter. The main results,
which stay qualitatively unchanged, are the following. For the center of
mass wave function of a homogeneous body of mass $M$ and linear dimensions $%
R $, effective gravitational self-interactions lead to a localization length 
$\Lambda \sim (\hslash ^{2}R^{3}/GM^{3})^{1/4}$, as soon as it is small with
respect to $R$. This produces a rather sharp threshold that, for ordinary
densities $\sim 10^{24}m_{p}/cm^{3}$, where $m_{p}$ denotes the proton mass,
is around $10^{11}m_{p}$, below which the effects of the effective
gravitational self-interactions are irrelevant\cite{defilippo1}. A localized
state slowly evolves, with times $\sim 10^{3}\sec $, for ordinary densities,
into a delocalized ensemble of localized states\cite{defilippo1}, this
entropic spreading replacing the wave function spreading of ordinary QM and
the unphysical stationary localized states of the nonlinear unitary
Schroedinger-Newton (S-N) model \cite{christian,penrose1,moroz,kumar}. As to
an unlocalized pure state of a body above threshold, it rapidly gets
localized, within times $\sim 10^{20}(M/m_{p})^{-5/3}\sec $, under
reasonable assumptions on the initial state, into such an ensemble\cite
{defilippo1}. This gives a well-defined dynamical model for gravity-induced
decoherence, to be compared with purely numerological estimates\cite
{karolyhazy,hawking3,diosi,ghirardi,penrose,squires,ellis2,anandan} and
which allows us to address physically relevant problems, like the
characterization of gravitationally decoherence free states of the physical
operator algebra\cite{defilippo0}. It is worthwhile to remark that, in spite
of the presence of the masses $\mu _{i},i=0,2$ (actually $\hslash c\mu _{i}$%
), the Newtonian limit has, for all practical purposes, no free parameter,
as to ordinary laboratory physics, if as usual they are assumed to be of the
order of the Planck mass, which is equivalent to take the limit $\mu
_{i}\rightarrow \infty $.

If the traditional Hamiltonian includes the Newton interaction, there are
extremely small violations of energy conservation, as only the
meta-Hamiltonian $H_{G}$ is strictly conserved. These fluctuations are
consistent with the assumption that an eigenstate of \ the traditional
Hamiltonian may evolve towards a microcanonical mixed state with an energy
dispersion around the original energy, which, though irrelevant on a
macroscopic scale, paves the way for the possibility that the thermodynamic
entropy of a closed system may be identified with its von Neumann entropy 
\cite{kay}. This is not irrelevant, if ''...in order to gain a better
understanding of the degrees of freedom responsible for black hole entropy,
it will be necessary to achieve a deeper understanding of the notion of
entropy itself. Even in flat space-time, there is far from universal
agreement as to the meaning of entropy -- particularly in quantum theory --
and as to the nature of the second law of thermodynamics''\cite{wald}. Of
course the reversibility of the unitary meta-dynamics makes entropy decrease
conceivable too\cite{schulman}, so that a derivation of the entropy-growth
for a closed system in the present context must have recourse to the choice
of suitable initial conditions, like unentanglement between the observable
and the hidden algebras\cite{kay}. While the assumption of special initial
conditions dates back to Boltzmann, only a non-unitary dynamics makes it a
viable starting point, within a quantum context, for the microscopic
derivation of the second law of thermodynamics, in terms of von Neumann
entropy, without renouncing strict isolation\cite{gemmer}.

Emboldened by the mentioned bonuses coming from the Newtonian limit of our
model, we now try to apply it to gravitational collapse. First evaluate
within our model the linear dimension of a collapsed matter lump, replacing
the classical singularity. In order to do that we boldly use Eq. (\ref
{metahamiltonian}) for lengths smaller than $\mu _{0}^{-1}$ and $\mu
_{2}^{-2}$, namely in the limit $\mu _{0},\mu _{2}\rightarrow 0$. This
corresponds to the replacement of our meta-Hamiltonian with the
meta-Hamiltonian in Ref. \cite{defilippo1}, where there is no gravitational
interaction within observable and within hidden matter, while there is a
Newton interaction between observable and hidden matter. This interaction is
effective in lowering the gravitational energy of a matter lump as far as
the localization length $\Lambda =(\hslash ^{2}R^{3}/GM^{3})^{1/4}$ is
fairly smaller than the lump radius $R$\cite{defilippo1}. The highest
possible density then corresponds roughly to $\Lambda =R$, namely to 
\begin{equation}
R=\hslash ^{2}/\left( GM^{3}\right) .  \label{minimal}
\end{equation}
In fact, below the localization threshold, only the interactions within
observable (and hidden meta-) matter are effective in collapsing matter,
but, in the considered limit $\mu _{0},\mu _{2}\rightarrow 0$, they vanish.
This parenthetically shows that the following discussion depends crucially,
not only on the doubling of the matter degrees of freedom, but also on the
inclusion of the repulsive interactions of HD gravity.

As to the space-time geometry, the Schwarzschild metric in ingoing
Eddington-Finkelstein coordinates ($v,r,\theta ,\phi $) covers the two
regions of the Kruskal maximal extension that are relevant to gravitational
collapses\cite{townsend}: 
\begin{equation}
ds^{2}=-\left[ 1-2MG/\left( rc^{2}\right) \right] dv^{2}+2drdv+r^{2}\left[
d\theta ^{2}+\sin ^{2}\theta d\phi ^{2}\right] .
\end{equation}
If, in the region beyond the horizon we put $x=v-\int dr\left[ 1-2MG/\left(
rc^{2}\right) \right] ^{-1}$, then 
\begin{equation}
ds^{2}=\left[ 1-2MG/\left( rc^{2}\right) \right] ^{-1}dr^{2}-\left[
1-2MG/\left( rc^{2}\right) \right] dx^{2}+r^{2}\left[ d\theta ^{2}+\sin
^{2}\theta d\phi ^{2}\right] ,  \label{xrmetric}
\end{equation}
where we see that beyond the horizon $r$ may be regarded as a time variable 
\cite{townsend}.

If we trust (\ref{minimal}) as the minimal length involved in the collapse,
we are led to assume that a full theory of quantum gravity should include a
mechanism regularizing the singularity at $r=0$ by means of that minimal
length. In particular, to characterize the region occupied by the collapsed
lump, consider that for time-like geodesics at constant $\theta $ and $\phi $
one can show that $\left| dx/dr\right| \sim r^{3/2}$ \ as $r\rightarrow 0$.
This implies that, the $x$ coordinate difference $\Delta x$\ of two material
points has a well defined limit as $r\rightarrow 0$, by which it is natural
to assume that the $x$ width of the collapsed matter lump is $\Delta x\sim R$%
. As to the apparent inconsistency of matter occupying just a finite $\Delta
x$ interval with $\partial /\partial x$ being a Killing vector, one should
expect on sub-Planckian scales substantial quantum corrections to the
Einstein equations that the model gives on a classical level (\ref{einstein}%
), with the dilaton and the ghost fields, though vanishing in the average,
playing a crucial role. On the other hand we are proceeding according to the
usual assumption, or fiction, of QM on the existence of a global time
variable, at least in the region swept by the lump. In fact the most natural
way to regularize (\ref{xrmetric}) is to consider it as an approximation for 
$r>R$ of a regular metric, whose coefficients for $r\rightarrow 0$
correspond to the ones in (\ref{xrmetric}) with $r=R$, in which case there
is no obstruction in extending the metric to $r<0$, where taking constant
coefficients makes $\partial /\partial r$ a time-like Killing vector. As a
consequence, the relevant space metric in the region swept by the collapsed
lump is 
\begin{equation}
ds_{SPACE}^{2}\sim 2MG/\left( Rc^{2}\right) dx^{2}+R^{2}\left[ d\theta
^{2}+\sin ^{2}\theta d\phi ^{2}\right] .
\end{equation}
The volume of the collapsed matter lump is then: 
\begin{equation}
V\sim R^{2}\Delta x\sqrt{MG/\left( Rc^{2}\right) }=\left[ \hslash
^{2}/\left( GM^{3}\right) \right] ^{5/2}\sqrt{MG/c^{2}}=\hslash
^{5}M^{-7}/\left( G^{2}c\right) .
\end{equation}
According to the above view, thermodynamical equilibrium is reached, due to
the gravitational interaction generating entanglement between the observable
and hidden meta-matter, by which the matter state is a microcanonical
ensemble corresponding to the energy 
\begin{equation}
E=Mc^{2}+GM^{2}/R=Mc^{2}+GM^{2}\left[ GM^{3}/\hslash ^{2}\right] \sim
G^{2}M^{5}/\hslash ^{2},\ \ if\ \ M\gg M_{P},
\end{equation}
where $M_{P}=\sqrt{\hslash c/G}$ is the Planck mass, and to the energy
density 
\begin{equation}
\varepsilon =E/V\sim G^{4}cM^{12}/\hslash ^{7}.
\end{equation}
We first treat the collapsed lump as a three-dimensional bulk in spite of
the huge dilation factor in the $x$ direction. Since this energy density
corresponds to a very high temperature, not to be mistaken for the Hawking
temperature, the matter can be represented by massless fields, whose
equilibrium entropy is given by 
\begin{equation}
S\sim \left( K_{B}/\not{h}^{3/4}c^{3/4}\right) \varepsilon
^{3/4}V=GM^{2}K_{B}/\left( \hslash c\right) .  \label{entropy}
\end{equation}
Of course this result can be trusted at most for its order of magnitude, the
uncertainty in the number of species being just one part of an unknown
numerical factor. With this proviso, common to other approaches\cite{wald},
Eq. (\ref{entropy}) agrees with B-H entropy. As this result contains four
dimensional constants, it is more than pure numerology: from a dimensional
viewpoint one could replace the dimensionless quantity $GM^{2}/\hslash c$ in
Eq. (\ref{entropy}) by any arbitrary function of $GM^{2}/\hslash c$. For
instance, if we ignored that the space geometry near the smoothed
singularity is not euclidean and then we assumed that $V$ $\sim R^{3}$, we
would get the entropy to be independent from the black hole mass, since the
increase of the entropy density with the mass would be offset by the
shrinking of the volume.

One could object, against the above derivation, that the collapsed matter
lump, for the presence of the dilation factor along the $x$ direction, is
more like a one-dimensional string-like structure of transverse dimensions $%
\sim $\ $R$ along $\theta $ and $\phi $ and length $L=(MG/Rc^{2})^{1/2}R\gg
R $. If we treat it like this, for the linear density of entropy we have $%
s\varpropto \varepsilon ^{1/2}$ and for the length $L\varpropto M^{-1}$, by
which $S=Ls\varpropto L(E/L)^{1/2}\varpropto M^{2}$, which agrees with the
previous result and with B-H entropy.

Finally, if we give for granted that a future theory of quantum gravity will
account for black hole evaporation, we can connect the temperature 
\begin{equation}
T\sim \sqrt[4]{\varepsilon h^{3}c^{3}}/K_{B}\sim cGM^{3}/K_{B}\hslash 
\end{equation}
of our collapsed matter lump with the temperature of the radiation at
infinity. If we model radiation by massless fields, emitted for simplicity
at a constant temperature as we are interested just in orders of magnitude,
this temperature is defined in terms of he ratio $E_{\infty }/S_{\infty }$
of its energy $E_{\infty }$ and its entropy $S_{\infty }$. It is natural to
assume that, ''once'' thermodynamical equilibrium is reached due to the
highly non-unitary dynamics close to the classical singularity, no entropy
production occurs during evaporation, by which $S_{\infty }=S$. Then, if $%
E_{\infty }=Mc^{2}$ is the energy of the total Hawking radiation spread over
a very large space volume, its temperature agrees with Hawking temperature,
i.e. \ 
\begin{equation}
T_{\infty }=\left( E_{\infty }/E\right) T\sim \left( c^{3}\hslash
/MGK_{B}\right) .
\end{equation}

Acknowledgments - I would like to thank Filippo Maimone, Luigi Mercaldo and
Mario Salerno for their warm encouragement and stimulating comments.
Financial support from M.U.R.S.T., Italy, I.N.F.M. and I.N.F.N., Salerno is
acknowledged.

\end{document}